\documentclass[preprint,review,12pt]{elsarticle}

\DeclareTextSymbol{\degre}{T1}{6}
\DeclareTextSymbol{\degre}{OT1}{23}

\usepackage{graphics}
\usepackage{amssymb}
\usepackage{color}
\usepackage{lineno}

\journal{Icarus}

\begin{document}

\begin{frontmatter}

%% Title, authors and addresses

%% use the tnoteref command within \title for footnotes;
%% use the tnotetext command for the associated footnote;
%% use the fnref command within \author or \address for footnotes;
%% use the fntext command for the associated footnote;
%% use the corref command within \author for corresponding author footnotes;
%% use the cortext command for the associated footnote;
%% use the ead command for the email address,
%% and the form \ead[url] for the home page:
%%
%% \title{Title\tnoteref{label1}}
%% \tnotetext[label1]{}
%% \author{Name\corref{cor1}\fnref{label2}}
%% \ead{email address}
%% \ead[url]{home page}
%% \fntext[label2]{}
%% \cortext[cor1]{}
%% \address{Address\fnref{label3}}
%% \fntext[label3]{}

%%\dochead{}
%% Use \dochead if there is an article header, e.g. \dochead{Short communication}
%% \dochead can also be used to include a conference title, if directed by the editors
%% e.g. \dochead{17th International Conference on Dynamical Processes in Excited States of Solids}

\title{Martian zeolites as a source of atmospheric methane}

\author{Olivier~Mousis}
\ead{olivier.mousis@lam.fr}
\address{Aix Marseille Universit\'e, CNRS, LAM (Laboratoire d'Astrophysique de Marseille) UMR 7326, 13388, Marseille, France}
\author{Jean-Marc Simon, Jean-Pierre Bellat}
\address{Laboratoire Interdisciplinaire Carnot de Bourgogne, UMR 6303, CNRS-Universit\'e de Bourgogne Franche Comt\'e, Dijon, France}
\author{Fr\'ed\'eric Schmidt, Sylvain Bouley, Eric Chassefi\`ere}
\address{Laboratoire GEOPS (G\'eosciences Paris Sud), Bat. 509, Universit\'e Paris Sud, 91405 Orsay Cedex, France}
\author{Violaine Sautter}
\address{Mus\'eum d'Histoire Naturelle, Paris, France}
\author{Yoann Quesnel}
\address{Aix-Marseille Universit\'e, CNRS, IRD, CEREGE UM34, 13545 Aix-en-Provence, France}
\author{Sylvain Picaud}
\address{Universit{\'e} de Franche-Comt{\'e}, Institut UTINAM, CNRS/INSU, UMR 6213, Besan\c con Cedex, France}
\author{S\'ebastien Lectez}
\address{Leeds University, School of Earth and Environment, Leeds, United Kingdom}

\begin{abstract}
The origin of the martian methane is still poorly understood. A plausible explanation is that methane could have been produced either by hydrothermal alteration of basaltic crust or by serpentinization of ultramafic rocks producing hydrogen and reducing crustal carbon into methane. Once formed, methane storage on Mars is commonly associated with the presence of hidden clathrate reservoirs. Here, we alternatively suggest that chabazite and clinoptilolite, which belong to the family of zeolites, may form a plausible storage reservoir of methane in the martian subsurface. Because of the existence of many volcanic terrains, zeolites are expected to be widespread on Mars and their Global Equivalent Layer may range up to more than $\sim$1 km, according to the most optimistic estimates. If the martian methane present in chabazite and clinoptilolite is directly sourced from an abiotic source in the subsurface, the destabilization of a localized layer of a few millimeters per year may be sufficient to explain the current observations. The sporadic release of methane from these zeolites requires that they also remained isolated from the atmosphere during its evolution. The methane release over the ages could be due to several mechanisms such as impacts, seismic activity or erosion. If the methane outgassing from excavated chabazite and/or clinoptilolite prevails on Mars, then the presence of these zeolites around Gale Crater could explain the variation of methane level observed by Mars Science Laboratory.
\end{abstract}

\begin{keyword}
%% keywords here, in the form: keyword \sep keyword

Mars \sep Mars, atmosphere \sep Mars, surface \sep Mineralogy \sep Astrobiology

%% PACS codes here, in the form: \PACS code \sep code

%% MSC codes here, in the form: \MSC code \sep code1
%% or \MSC[2008] code \sep code (2000 is the default)

\end{keyword}

\end{frontmatter}

%%
%% Start line numbering here if you want
%%
% \linenumbers

%% main text
\section{Introduction}

The origin of the martian methane (CH$_4$) is still poorly understood. Despite the fact that the presence of CH$_4$ remains under debate (Zahnle et al. 2011; Zahnle 2015), detections have been claimed at the 10--60 parts per billion by volume (ppbv) level in Mars' atmosphere from space and ground-based observations at the end of the 90s and during the following decade (Formisano et al., 2004; Krasnopolsky et al. 2004; Mumma et al. 2009; Fonti and Marzo 2010). Recent observations suggest a CH$_4$ atmospheric abundance of $\sim$10 ppbv, and in some cases no or little CH$_4$ with an upper limit of $\sim$7 ppbv in 2009--2010, during Mars' northern spring (Krasnopolsky 2012; Villanueva et al. 2013). More recent {\it in situ} measurements performed by Mars Science Laboratory (MSL) have evidenced variations in the methane detection at the location of Gale Crater. Despite a background level of methane remaining at 0.69~$\pm$~0.25 ppbv, an elevated level of methane of 7.2~$\pm$~2.1 ppbv was evidenced during a timespan of $\sim$6 months (see Table 1 of Webster et al. 2015), a range of values comparable to the levels observed remotely during the last decade.

Because local methane enhancements such as those measured by MSL require CH$_4$ atmospheric lifetimes of less than 1 yr (Lef\`evre and Forget 2009), its release from a subsurface reservoir or an active primary source has widely been discussed in the literature. A plausible explanation is that CH$_4$ could have been produced either by hydrothermal alteration of basaltic crust (Lyons et al. 2005) or by serpentinization of ultramafic rocks producing H$_2$ and reducing crustal carbon into CH$_4$ (Oze and Sharma 2005; Atreya et al. 2007; Chassefi\`ere and Leblanc 2011; Chassefi\`ere et al. 2013; Holm et al. 2015). This hypothesis is supported by the fact that ultramafic and serpentinized rocks have been observed on Mars, in particular in the Nili Fossae region (Brown et al. 2010; Ehlmann et al. 2010; Viviano et al. 2013). Once formed, methane storage on Mars is commonly associated with the presence of hidden clathrate reservoirs. Martian clathrates would form an intermediate storage reservoir in the subsurface that regularly releases methane into the atmosphere (Prieto-Ballesteros et al. 2006; Chastain and Chevrier 2007; Thomas et al. 2009; Gainey and Elwood Madden 2012; Herri and Chassefi\`ere 2012; Mousis et al. 2013, 2015). However, because clathrates are more likely thermodynamically stable in the martian subsurface and at depths depending on the soil's porosity (Mousis et al. 2013), their existence has never be proven by remote or {\it in situ} observations. Interestingly, it has been recently proposed that halite or regolith could also sequestrate CH$_4$ on the martian surface (Fries et al. 2015; Hu et al. 2015), but these mechanisms still need to be thoroughly investigated.

Here, because of their ability to trap substantial amounts of gases, we suggest that zeolites may form an alternative plausible storage reservoir of methane in the martian subsurface. Spectral evidence for the presence of zeolite has been found on the martian surface (Ruff 2004; Ehlmann et al. 2009; Carter et al. 2013; Ehlmann 2014) and there is strong geological case arguing for the presence of this aluminosilicate as part of the martian regolith. In Sec. 2, we explain why chabazite, analcime and clinoptilolite are good candidates to account for the widespread occurrence of zeolites on Mars. We also provide an estimate of the amount of zeolites potentially existing on the planet. Sec. 3 is dedicated to the description of the adsorption properties of chabazite, analcime and clinoptilolite. The amount of methane potentially trapped in these zeolites in martian conditions is estimated in Sec. 4. Sec. 5 is devoted to discussion.

\section{Zeolites on Mars}

Zeolites have been first detected by Ruff (2004) on martian dust using the Mars Global Surveyor (MGS) TES spectroscopic observations. Fialips et al. (2005) then suggested that the water-equivalent hydrogen observed by Mars Odyssey could be partially stored by zeolite minerals present in the first meters in the martian regolith. Indeed, Dickinson and Rosen (2003) observed up to 18 wt\% of authigenic chabazite in frozen soils of Antarctica (equivalent to martian conditions). Recently, both OMEGA and CRISM instruments onboard the ESA Mars Express and NASA Mars Reconnaissance Orbiter (MRO) detected zeolite minerals on the rocky outcrops of several places on Mars (Ehlmann et al. 2009; Carter et al. 2013, Ehlmann 2014). While the first observations on dust and soils suggested a grossly zeolite mineral distribution at mid-latitude, we now have detailed observations revealing the geological/morphological context of zeolite outcrops (152 occurrences were detected by Carter et al. 2013). For instance, Ehlmann et al. (2009) claimed the identification of pure analcime (Si-Al-Na form) in the deposits in and around the central peaks of two 25-km impact craters nearby Nili Fossae and Isidis. These peaks would then reflect post-impact hydrothermal alteration (Osinski and Pierazzo 2013). Carter et al. (2013) had a detailed discussion of the issue of their timing of formation and concluded that most hydrous minerals, including zeolites, were formed during the Noachian period. However, they also noticed the presence of zeolites in the younger northern lowlands, probably resulting from ice-volcano interaction.
   
In summary, both TES and OMEGA instruments were able to remotely differentiate zeolite spectra from other alteration minerals, namely opal A and saponite formed under similar conditions. Such secondary zeolites result from low temperature aqueous alteration by alkaline brines (or ice) of volcanic glass included in pyroclastic or volcanic sedimentary rocks and form authigenic cements in volcanoclastic sandstone. Note that volcanic ash and tephra, the common contributor to sedimentary material on Mars, should be widespread, as explosive volcanism on Mars is the rule rather than the exception (Grott et al. 2013). However, the resolution of existing infrared spectra remains insufficient to constrain the variety of zeolites that really crystallized on Mars. 

Among the possible zeolites, chabazite is a good candidate to account for their widespread occurrence on Mars. This mineral is the end product of weathering sequences in a wide range of chemical context ranging from silica-rich to silica-poor volcanic rocks. Chabazite typically forms in chemically open systems, in which transports of soluble ions take place efficiently by flowing vadose water or near-surface ground water (Sheppard and Hay 2001). On the other hand, in the closed systems in the martian subsurface, more alkali analcime and clinoptilolite should be the major zeolites due to limitation of transports of soluble ions. Also, there are several terrestrial locations where nearly pure analcimes form thick bedding (several tens of meters) with wide special extent (hundreds of kilometers) (Sheppard and Gude 1973; Whateley et al. 1996; Deer et al. 2004).

One can provide an estimate of the amount of zeolites potentially existing on Mars. Using Noachian estimates for the martian crustal thermal flux (12--20$\degre$C/km) and thermodynamic data of low-grade metamorphic facies ($\sim$160--220 $\degre$C and from 0 to 3 $\times$10$^5$ kPa), zeolites may be formed at depths ranging from approximately 8 to 15--20 km (e.g. McSween et al. 2015). This estimate is confirmed by the detection of zeolites near central peaks, independently suggesting that those minerals are indeed present at depths of several kilometers in the crust. Assuming this depth range (8--15 km), it corresponds to a global volume of 10$^9$ km$^3$ and a Global Equivalent Layer (GEL) reaching $\sim$7 km of martian zeolites. However, the maps of Carter et al. (2013) show that the area where zeolites (and all hydrous minerals) were detected by remote sensing is equivalent to the surface of the 0--45$\degre$S latitudinal band, i.e. about 35\% of the surface of Mars. If we do not consider temperature constraints but only different thicknesses (0.001 to 10 km) of a 100\% zeolite layer at all 0--45 $\degre$S latitudes, the total volume and GEL are in the $\sim$5 $\times$ 10$^4$--5 $\times$ 10$^8$ km$^3$ and $\sim$0.35--3500 m ranges, respectively. The smallest values may be considered as reasonable estimates (in the range of $\sim$1\% of zeolite in crystal clays; Ehlmann et al. 2011), but other geological settings or models can be considered. For instance, the total volume of possible isolated cylindrical zeolite layers located beneath $\sim$150 impact craters (Carter et al. 2013) ranging from 5 to 200 km of diameter (zeolite layer thickness from 0.1 km to unrealistic 20 km) may reach $\sim$10$^3$ to 10$^8$ km$^3$ (0.07 to 700 m GEL). These isolated layers may correspond to zeolite minerals formed by post-impact hydrothermal alteration (Osinski and Pierazzo 2013). Therefore, it seems that any scenario of zeolite geological generation (sparse post-impact hydrothermal alteration or crustal global alteration) can lead to important ranges of volumes/GEL. These values, in particular the most optimistic ones, should not be taken as true quantities, but only as starting reasonable estimates. Indeed crustal porosity and fluids surely decrease the efficiency of zeolite formation at depths : the low water-to-rock ratio would prevent any alteration while secondary mineralizations fill the pores.

\section{Adsorption properties of zeolites}

In this Section, the adsorption selectivity of CH$_4$ with respect to CO$_2$ is investigated on chabazite, analcime and clinoptilolite.

\subsection{Chabazite}

\begin{figure}
\begin{center}
\resizebox{\hsize}{!}{\includegraphics[angle=0]{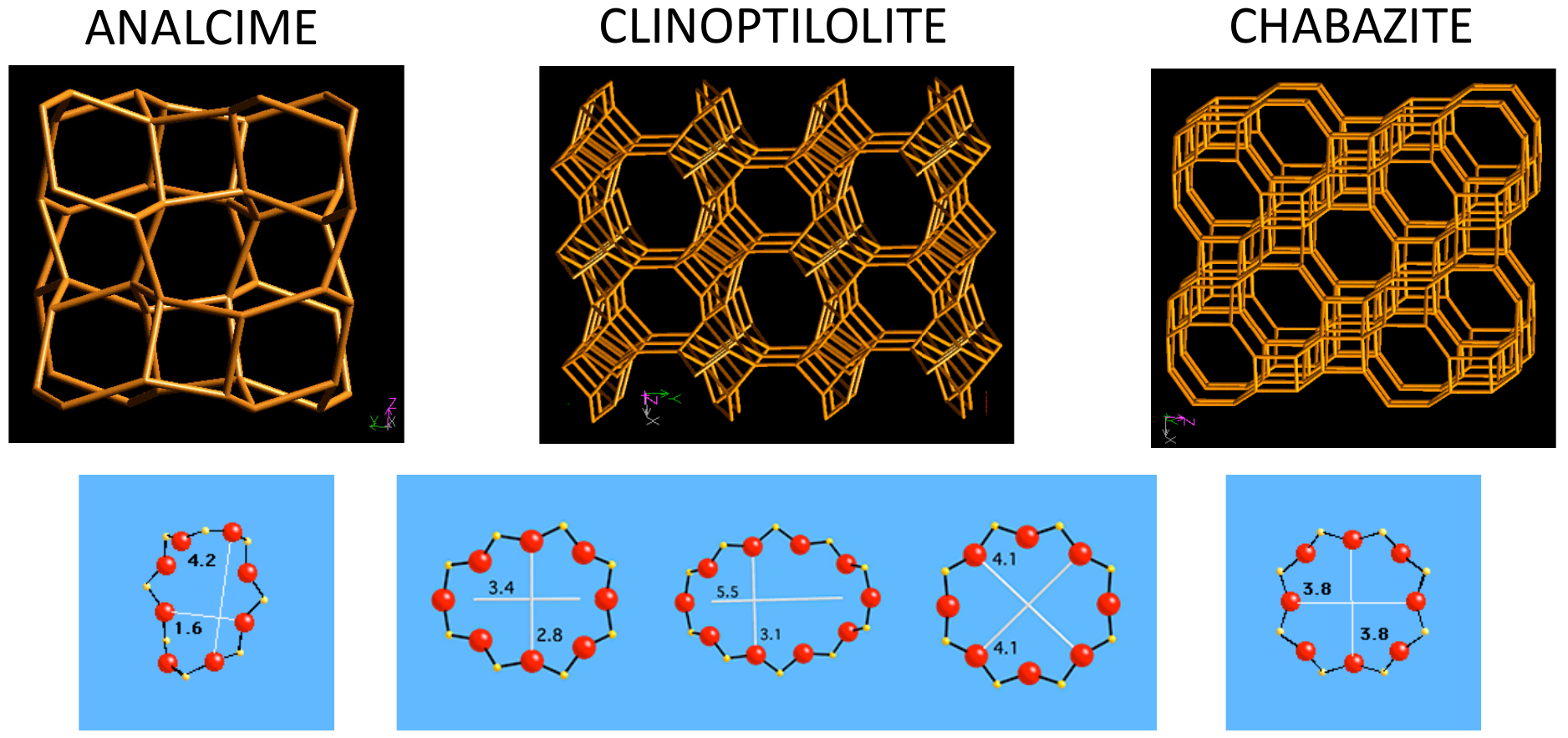}}
\caption{Framework type of zeolites. Dimensions of pore openings are given in~\AA. The maximum diameter of molecules that can diffuse in these frameworks are around 2.43~\AA~for analcime, 3.67~\AA~for clinoptilolite and 3.72~\AA~for chabazite. (From Database of Zeolite Structures http://www.iza-structure.org).}
\label{chab1}
\end{center}
\end{figure}

The common chemical formula of a hydrated chabazite is [Ca$_6$Al$_{12}$Si$_{24}$O$_{72}$],(H$_2$O)$_{40}$. There exists several forms of chabazite zeolites that differ in their Si/Al ratio and the nature of cations (Ba$^{2+}$, Ca$^{2+}$, Sr$^{2+}$, K$^+$, Na$^+$), which counterbalance the electric charges. The framework structure of chabazite is composed of SiO$_4$ and AlO$_4$ tetrahedrons joined by their oxygen atoms. This arrangement forms primary building units interconnected by secondary building units, as shown in Fig. \ref{chab1}. The unit cell of chabazite thus contains one large ellipsoidal cavity accessed by six 8-ring windows (Pascale et al. 2002). 

\begin{figure}
\begin{center}
\resizebox{\hsize}{!}{\includegraphics[angle=0]{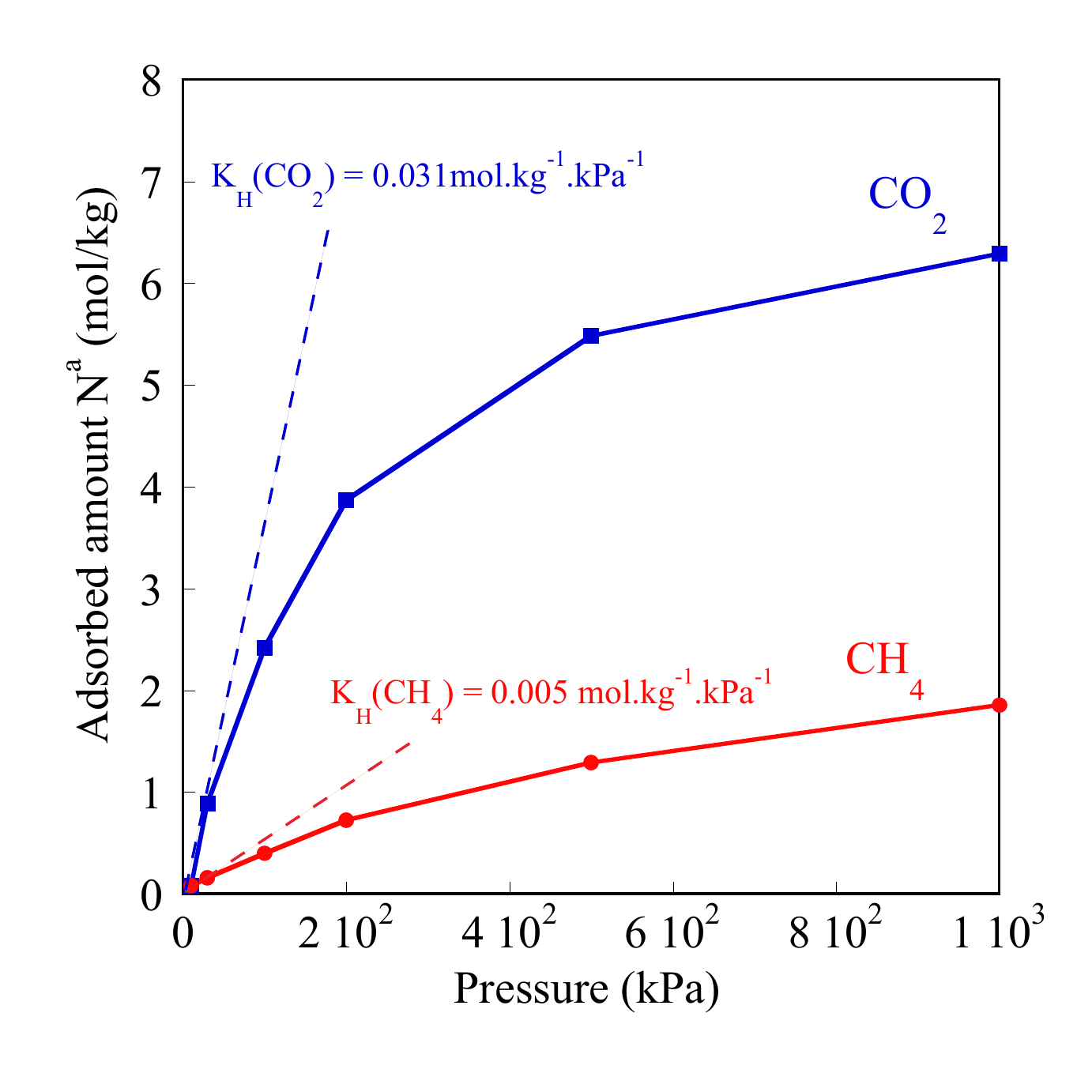}}
\caption{Single adsorption isotherms of CO$_2$ and CH$_4$ on chabazite as a function of pressure at 300 K. Symbols: GCMC simulations and experiments (Garcia-Perez et al. 2007). Dashed lines: Henry's law region.}
\label{chab2}
\end{center}
\end{figure}

The chabazite zeolite gets specific adsorption properties for various molecules having a size smaller than the 8-ring apertures, which gives them access to the ellipsoidal cages. Owing to the presence of compensation alkali cations, the chabazite is a hydrophilic material. Its adsorption capacity of water is around 0.2 cm$^3$/g at 257 K (J\"anchen et al. 2006). This zeolite is also able to adsorb CO$_2$ and CH$_4$, two molecules of interest for the martian atmosphere. Figure \ref{chab2} shows the adsorption isotherms of CO$_2$ and CH$_4$ on chabazite at 300 K calculated by Monte Carlo simulations in the grand canonical ensemble (GCMC) (Garcia-Perez et al. 2007) for pressure ranging below 10$^3$ kPa. These simulations are in a very good agreement with some experimental data reported in the literature (Watson et al., 2012; Jensen et al. 2012). 

As would expected, owing to the presence of a quadripolar moment in the carbon dioxide molecule, the chabazite adsorbs more CO$_2$ than CH$_4$. In spite of this, the amount of methane adsorbed is not insignificant at 300 K and even better at lower temperature. It can reach $\sim$2 mol/kg against more than 6 mol/kg in the case of CO$_2$. This result suggests that the chabazite will selectively adsorb methane and carbon dioxide, with an adsorption in favor of the latter molecule. The adsorption selectivity of methane with respect to carbon dioxide is defined by the relation:

\begin{equation}
\alpha_{\rm CH_4/CO_2} = \frac{x_{\rm CH_4}/y_{\rm CH_4}}{x_{\rm CO_2}/y_{\rm CO_2}},
\label{eq1}
\end{equation}

\noindent where $x_{\rm i}$ and $y_{\rm i}$ are the mole fractions of component i in the adsorbed phase and in the gas phase at equilibrium, respectively. $\alpha_{\rm CH_4/\rm CO_2}$ can be predicted from the adsorption isotherms of single components by means of the ideal adsorbed solution theory (IAS theory; Myers and Prausnitz. 1965). When the gas pressure converges towards zero, each single adsorption isotherm exhibits a linear part (see Fig. \ref{chab2}), which corresponds to the Henry's law region. In this domain, the adsorbed amount of each single component i is proportional to the gas pressure: $N^a_{\rm i}$~=~$K_{H, \rm i}~P$. By applying the IAS theory to the Henry's law region, this allows us to derive the adsorption selectivity from the ratio between the Henry constants:

\begin{equation}
\alpha_{\rm CH_4/CO_2} = \frac{K_H({\rm CH_4})}{K_H({\rm CO_2})}.
\label{eq2}
\end{equation}

The ratio of the adsorbed amounts of CH$_4$ and CO$_2$ can then be related to the ratio of their partial pressures at equilibrium via the relation:

\begin{equation}
\frac{N^a_{\rm CH_4}}{N^a_{\rm CO_2}} = \alpha_{\rm CH_4/CO_2} \frac{P_{\rm CH_4}}{P_{\rm CO_2}}.
\label{eq3}
\end{equation}

\noindent Henry constants determined from experimental adsorption isotherms at 300 K are given in Table \ref{table_chab} with the corresponding selectivities. With the values of the adsorption enthalpies found in the literature, the Henry constants can be estimated at any temperature relevant to Mars' conditions by using the van't Hoff relation:

\begin{equation}
\frac{d}{dT} ln(K_{H, \rm i}) = \frac{\Delta H}{RT^2}.
\label{eq4}
\end{equation}

\noindent Once the values of $K_{H, \rm i}$ have been determined for CH$_4$ and CO$_2$ at given temperature, it is possible to derive the corresponding adsorption selectivity from Eq. \ref{eq2}.

\subsection{Analcime}

The theoretical chemical formula of analcime is [NaAlSi$_2$O$_6$],(H$_2$O). The structure of analcime, represented in Fig. \ref{chab1}, is very constricted; the basic SiO$_4$ and AlO$_4$ tetrahedra mutually link to form 4 or 6 membered rings. The maximum diameter of a sphere that can diffuse throughout this structure is $\sim$2.4 \AA. Because the kinetic diameters of CO$_2$ and CH$_4$ are 3.3 and 3.8 \AA~respectively, these two molecules cannot be adsorbed in analcime. Only water can be adsorbed in analcime, due to its smaller diameter ($\sim$2.6 \AA).

\subsection{Clinoptilolite}

The common chemical formula of clinoptilolite is [M$_6$Al$_6$Si$_{30}$O$_{72}$],(H$_2$O)$_{12}$, where M is a compensation cation easily exchangeable which can be Na, K, Ca, Sr, Ba and Mg according to the source of minerals (Sand et al. 1978). Clinoptilolite has the same framework structure as heulandite. However it presents a better thermal stability. The porosity is composed of three sets of intersecting channels, all in the same plane (Fig. 1): A channels with 8-membered rings (aperture 3 $\times$ 7.6 \AA), B channels parallel to A channels with 8-membered rings (aperture 3.3 $\times$ 7.6 \AA) and C channels quasi perpendicular to the two others with 8-membered rings (aperture 2.6 $\times$ 4.7 \AA) (Baerlocher et al. 2007). The microporous volume determined by water adsorption is around 0.16 cm$^3$/g at 298 K. Owing to the presence of compensation cations, clinoptilolite exhibits good adsorption affinity towards water, carbon dioxide and methane which, unike analcime, can enter its micropores despite a pore aperture close to the kinetic diameter of these molecules. As chabazite, this zeolite preferentially adsorbs carbon dioxide compared to methane. At 298 K, some varieties of clinoptilolite can adsorb more than 3.6 mol/kg of CO$_2$ at room temperature under 10 kPa (Breck et al. 1974) while only 0.25 mol/kg of CH$_4$ under the same conditions (Kouvelos et al. 2007). Here, the determination of the adsorption selectivity of CH$_4$ with respect to CO$_2$ in clinoptilolite is calculated following the same approach as for chabazite. Adsorption capacities, adsorption enthalpies, Henry constants and adsorption selectivities for CO$_2$ and CH$_4$ have been derived from the experiments of Arefi Pour et al. (2015).

\section{Methane trapping at low pressure regime}

In Sec. 3, we have shown that the application of the Henry's law allows to extrapolate the amounts of CH$_4$ and CO$_2$ trapped in chabazite or clinoptilolite at low pressure range. Because the current martian surface atmospheric pressure (0.6 kPa) is located in the validity domain of Henry's law (see the example of Fig. \ref{chab2} given at 300 K), this enables us to investigate the amount of CH$_4$ that would be potentially trapped in Martian chabazite or clinoptilolite in contact with an older martian atmosphere at various temperatures, assuming that the methane abundance was higher than today's value at that time.

\begin{figure}
\begin{center}
\resizebox{\hsize}{!}{\includegraphics[angle=0]{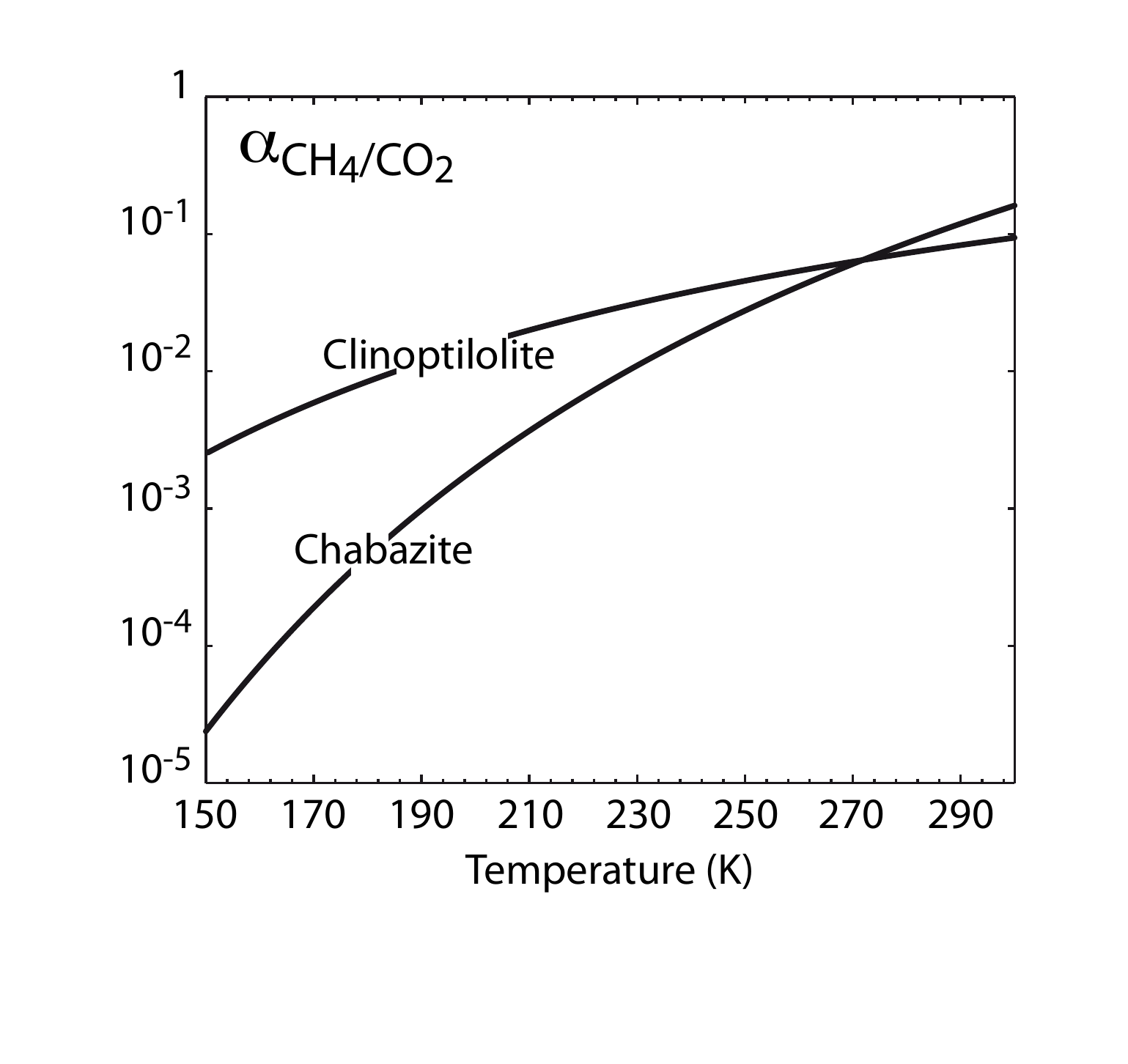}}
\caption{Adsorption selectivity of CH$_4$ with respect to CO$_2$ in chabazite and clinoptilolite as a function of temperature.}
\label{chab3}
\end{center}
\end{figure}

The adsorption selectivity of methane with respect to carbon dioxide $\alpha_{\rm CH_4/\rm CO_2}$ represents the ratio of the CH$_4$ abundance in chabazite or clinoptilolite to its abundance in the coexisting gas phase at low pressure range. The evolution of $\alpha_{\rm CH_4/\rm CO_2}$ as a function of temperature is illustrated by Fig. \ref{chab3} in the cases of the two zeolites. With values between 2.5~$\times$~10$^{-5}$ and 0.169 (chabazite) and between 2.5~$\times$~10$^{-3}$ and 0.094 (clinoptilolite) in the 150--300 K range, we find that the CH$_4$/CO$_2$ ratio increases with higher temperatures in both zeolites, regardless of the initial CH$_4$-CO$_2$ gaseous mixture.

\begin{figure}
\begin{center}
\includegraphics[width=9cm,angle=0]{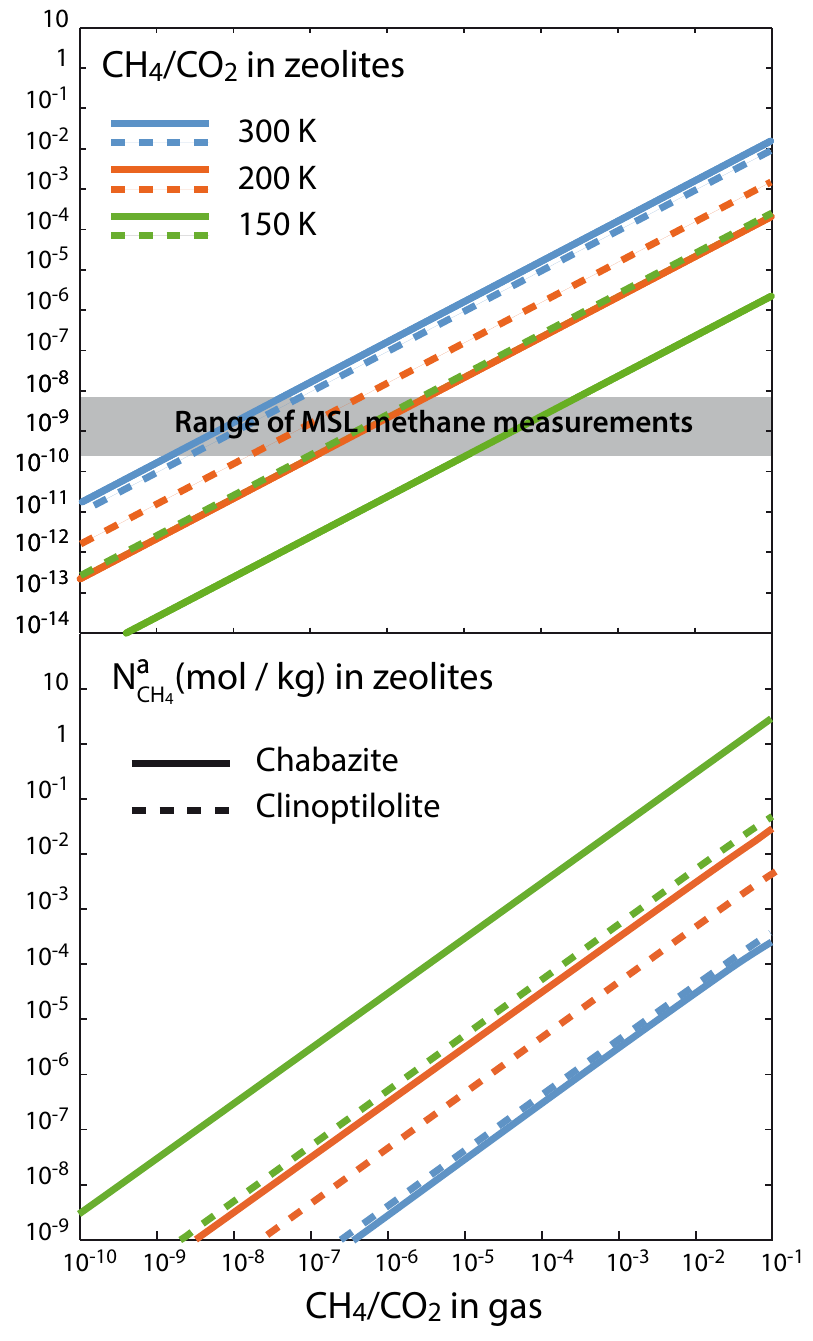}
\caption{Top: CH$_4$/CO$_2$ ratio in chabazite and clinoptilolite as a function of its ratio in coexisting gas represented at $T$ = 150, 200 and 300 K and 0.6 KPa of total pressure. The grey area corresponds to the range of CH$_4$ measurements made so far by MSL in the martian atmosphere. Bottom: amount of CH$_4$ trapped chabazite and clinoptilolite as a function of the CH$_4$/CO$_2$ ratio in coexisting gas calculated at $T$ = 150, 200 and 300 K and 0.6 KPa of total pressure.}
\label{chab4}
\end{center}
\end{figure}

Figure \ref{chab4} represents the evolution of the CH$_4$/CO$_2$ ratio in the two zeolites as a function of the CH$_4$/CO$_2$ ratio in the coexisting gas at three temperatures of interest, namely  the coldest winter temperature reached in the south pole region (150 K), and the average night (200 K) and day (300 K) surface temperatures at mid-latitudes. It shows that the CH$_4$/CO$_2$ ratio must be in the $\sim$10$^{-5}$--3~$\times$~10$^{-4}$ range at 150 K in the coexisting gas phase to give a value in chabazite matching the CH$_4$ abundance range measured by MSL (in the $\sim$0.25--7.2 $\times$ 10$^{-9}$ range). CH$_4$/CO$_2$ ratios must also exceed $\sim$5~$\times$~10$^{-8}$ and 4~$\times$~10$^{-6}$ in gas to give values in chabazite higher than those measured by MSL at 300 and 200 K, respectively. On the other hand, the CH$_4$/CO$_2$ ratio in the coexisting gas must be in the $\sim$10$^{-7}$--3~$\times$10$^{-6}$ range in clinoptilolite to match the CH$_4$ abundance range measured by MSL at 150 K. Interestingly, because $\alpha_{\rm CH_4/\rm CO_2}$ is higher in clinoptilolite than in chabazite at temperatures lower than $\sim$270 K, smaller CH$_4$/CO$_2$ ratios ($\sim$7~$\times$~10$^{-8}$ at 300 K and $\sim$5~$\times$~10$^{-7}$ at 200 K) are needed to allow this zeolite to match the MSL values.

%Figure \ref{chab4} also displays the amount of CH$_4$ trapped in chabazite as a function of the CH$_4$/CO$_2$ ratio in the coexisting gas at 150, 200 and 300 K. It shows that, despite a lower value of the adsorption selectivity $\alpha_{\rm CH_4/\rm CO_2}$, the amount of trapped methane is $\sim$4 orders of magnitudes higher at 150 K than at 300 K.}

\section{Discussion}

Comparisons with models depicting the composition of clathrates potentially existing in the martian subsurface show that chabazite or clinoptilolite can be comparable methane sinks (i.e., methane trapping from a methane-containing atmosphere on early Mars or from an abiotic source in the crust on early/present Mars) at significantly higher temperatures. For example, $\alpha_{\rm CH_4/CO_2}$ in chabazite or clinoptilolite becomes greater or equal to 0.1 at temperatures reaching 300 K (see Fig. \ref{chab3}), whatever the initial CH$_4$ atmospheric mole fraction. Similar selectivities are achieved in clathrates for CH$_4$ mole fractions in the 10$^{-4}$--10$^{-2}$ range but the existence of these structures requires temperatures lower than $\sim$150~K at 0.6 kPa of atmospheric pressure (Mousis et al. 2013), namely the coldest temperature reached during winter in the south pole region. Therefore, scenarios advocating a substantial trapping of volatiles in martian clathrates argue that these ices are buried in the soil at sufficient depth, allowing them to be isolated from the atmosphere and remain stable over long time periods (Chastain and Chevrier 2007; Thomas et al. 2009; Herri and Chassefi\`ere 2012; Chassefi\`ere et al. 2013; Mousis et al. 2013). This scenario also applies to martian chabazite or clinoptilolite, allowing CH$_4$ to be extracted either from a potentially methane-rich ancient atmosphere or directly from an abiotic source localized in the crust. Indeed, because of their burial in the soil, these zeolites could have preserved the trapped methane over long time periods and create the sporadic releases observed in the atmosphere over the last decade due to impacts, seismic activity or erosion. An alternative possibility would be to assume that zeolites did continuously remain in equilibrium with the martian atmosphere during the course of its evolution. In this case, zeolites would not be able to supply any methane to the atmosphere: because the value of $\alpha_{\rm CH_4/CO_2}$ is in the $\sim$10$^{-3}$--10$^{-1}$ range between 200 and 300 K, the amount of CH$_4$ trapped in chabazite and clinoptilolite would be lower than the measured atmospheric levels. For the same reason, chabazite and clinoptilolite could not act as CH$_4$ sinks if they remain in contact with the atmosphere. Similarly to the proposed trapping scenarios in clathrates, the methane stored in these zeolites could have been produced earlier either via hydrothermal alteration of basaltic crust (Lyons et al. 2005) or via serpentinization reactions (Oze and Sharma 2005; Atreya et al. 2007; Chassefi\`ere and Leblanc 2011; Chassefi\`ere et al. 2013; Holm et al. 2015). Otherwise, alternative methane sinks should be considered on Mars.

In order to quantitatively test the link between the current time presence of CH$_4$ in the atmosphere and the possible destabilization of zeolites, we can estimate the total GEL of zeolites that must be destabilized each second, assuming an initial quantity of trapped CH$_4$. In the following, we make the assumption that chabazite or analcime are the dominant zeolites on Mars. If, instead, analcime is the dominant form, then this material cannot be at the origin of the atmospheric CH$_4$, due to the small size of its porous network. Two case studies can be envisaged. In the first case, we assume that CH$_4$ is trapped in chabazite or clinoptilolite in contact with an ancient martian atmosphere at a surface pressure and temperature of 0.6 kPa and 150 K, respectively. Matching the upper MSL value at 150 K requires a CH$_4$/CO$_2$ ratio of 3 $\times$ 10$^{-4}$ and 3 $\times$ 10$^{-6}$ in the gas phase released by chabazite and clinoptilolite, respectively. These ratios correspond to amounts of trapped CH$_4$ of $\sim$10$^{-2}$ mol kg$^{-1}$ in chabazite and $\sim$10$^{-6}$ mol kg$^{-1}$ in clinoptilolite (see Fig. \ref{chab4}). The total injection flux of CH$_4$ in the Martian atmosphere has been estimated to be 85--100 kg s$^{-1}$ (Mischna et al. 2011; Holmes et al. 2015). Assuming a mean density of 2000 kg m$^{-3}$ for chabazite and clinoptilolite, this flux corresponds to at least $\sim$5.3~$\times$~10$^{7 \pm 2}$ kg s$^{-1}$ of zeolites and 5~$\times$~10$^{-3 \pm 2}$ m~yr$^{-1}$ as GEL. A more precise calculation would be based on the realistic value of localized surface flux 10$^{-11}$--10$^{-9}$ kg m$^{-2}$ s$^{-1}$ derived by Holmes et al. (2015), assuming a source within a homogeneous 5$\degre$x5$\degre$ region. In this case, the corresponding mass of destabilized zeolites would be 6~$\times$~10$^{-7 \pm 1}$ kg m$^{-2}$ s$^{-1}$ (chabazite) and  6~$\times$~10$^{-3 \pm 1}$ kg m$^{-2}$ s$^{-1}$ (clinoptilolite). These values correspond to localized layers with   thicknesses of $\sim$10$^{-3}$--10$^{-1}$ m for chabazite and $\sim$10--1000 m for clinoptilolite.

In the second case, we assume that chabazite or clinoptilolite are directly filled by pure methane produced from an abiotic source localized at depth in the crust. Here, given the high lithostatic pressure, the Henry's law does not apply anymore and single adsorption temperature isotherms of CH$_4$ derived from experiments or GCMC computations must be used. As a toy example, we consider the Figure \ref{chab2} which shows that at 300 K and 10$^3$ kPa of gas pressure (corresponding to a depth of $\sim$90 m), the amount of CH$_4$ trapped in chabazite reaches $\sim$2 mol kg$^{-1}$. Assuming a total CH$_4$ injection flux of 85--100 kg s$^{-1}$ in the Martian atmosphere, we find that it corresponds at least to $\sim$2650 kg s$^{-1}$ of chabazite and 2.9~$\times$~10$^{-7}$ m yr$^{-1}$ as GEL. Using the localized surface flux 10$^{-11}$--10$^{-9}$ kg m$^{-2}$ s$^{-1}$ sourced from a 5$\degre$x5$\degre$ region, the corresponding mass of destabilized chabazite would be 3~$\times$~10$^{-9 \pm 1}$ kg m$^{-2}$ s$^{-1}$, representing a localized layer of $\sim$5 $\times$ 10$^{-6}$--5 $\times$ 10$^{-4}$ m yr$^{-1}$. In the case of clinoptilolite, assuming an amount of trapped CH$_4$ 10 times smaller than the one estimated for chabazite (see Sec. 3.3), the aforementioned values of GEL and localized layer would be increased by the same factor.

 When based on realistic values of the methane flux, our calculations show that the second case appears more plausible than the first because it requires amounts of chabazite or clinoptilolite well below those independently quantified from geological constraints. If the martian methane present in chabazite or clinoptilolite is directly sourced from an abiotic source in the subsurface, the destabilization of a zeolite localized layer of a few millimeters per year at worst may be sufficient to explain the current observations. Our study suggests that if the methane outgassing from excavated chabazite or clinoptilolite prevails over any other source on Mars, then the presence of these minerals around Gale Crater could explain the variation of the CH$_4$ level observed by MSL. An interesting follow-up of this work would be to investigate the adsorption/desorption efficiencies of other gases in zeolites. Coupling all the data together might lead to predictions of other observable effects in the martian atmosphere that could be used to test the present hypothesis.

Finally, it is interesting to note that the zeolites adsorption properties depend on their Si/Al ratios. The smaller is this ratio, the greater are the compensating cations and the hydrophilic and organophilic properties of the zeolites. The adsorption selectivity of CH$_4$ with respect to CO$_2$ also varies according to this ratio and the nature of the cation. However, in the cases of chabazite and clinoptilolite, there will always be a preferential adsorption of CO$_2$ at the expense of CH$_4$, whatever the Si/Al ratio and the nature of the cation. This preferential adsorption results from the specific interactions induced by the quadrupole moment caused by the presence of pi electrons in CO$_2$, which is not the case with CH$_4$ which puts into action non-specific interactions.

\section*{Acknowledgments}

The work contributed by O.M. was carried out thanks to the support of the A*MIDEX project (n\textsuperscript{o} ANR-11-IDEX-0001-02) funded by the ``Investissements d'Avenir'' French Government program, managed by the French National Research Agency (ANR). We acknowledge support from the ``Observatoire des Sciences de l'Univers'' (OSU) Institut Pyth\'eas, the ``Institut National des Sciences de l'Univers''  (INSU), the ``Centre National de la Recherche Scientifique'' (CNRS) and ``Centre National d'Etude Spatiale'' (CNES) and through the ``Programme National de Plan\'etologie'' and MEX/PFS Program. We also thank two anonymous reviewers who made very interesting suggestions that helped us improve our first manuscript.

\clearpage

\begin{table*}[h]
\centering\caption{Adsorption enthalpies and Henry constants obtained from adsorption experiments of single components on chabazite and of binary mixtures on clinoptilolite.}
\begin{center}
\begin{tabular}{lcccccc}
\hline
& 											&\multicolumn{3}{c}{Chabazite} 						&\multicolumn{2}{c}{Clinoptilolite}	\\
 								&			&  150 K						& 200 K 	& 300 K		&  150 K		& 300 K			\\
\hline
$\Delta$H$$ (kJ mol$^{-1}$)			& CH$_4$		& \multicolumn{3}{c}{-23$^{(a)}$}						& \multicolumn{2}{c}{-12$^{(b)}$}	\\
    								& CO$_2$		& \multicolumn{3}{c}{-45$^{(a)}$}						& \multicolumn{2}{c}{-21$^{(b)}$}	\\
K$_H$ (mol kg$^{-1}$ kPa$^{-1}$)		& CH$_4$	 	& 50 							& 0.503 	& 0.005		& 0.877 		& 	0.007		\\	
 								& CO$_2$	 	& 2 $\times$ 10$^6$   			& 256 	& 0.031		& 345 		& 0.076			\\
\hline 
\end{tabular}
$^{(a)}$Jensen et al. (2012); $^{(a)}$Arefi Pour et al. (2015)
\end{center}
\label{table_chab}
\end{table*}

\clearpage

\clearpage

\clearpage

\clearpage

\end{document}